\documentclass[aps,prapplied,reprint]{revtex4-2}

\usepackage[mathlines]{lineno}
\usepackage{lipsum}
\usepackage{blindtext}
\usepackage[export]{adjustbox}
\usepackage{graphicx,epsfig,epstopdf}
\usepackage{amsmath,amssymb}
\usepackage{esvect}
\usepackage{mathtools}
\usepackage{color}
\usepackage{subfigure}
\usepackage{array}
\usepackage{tabularx}
\usepackage{multirow}
\usepackage{booktabs}
\usepackage{mathtools}
\usepackage{float}
\usepackage{hyperref}
\newcolumntype{L}[1]{>{\raggedright\arraybackslash}p{#1}}
\newcolumntype{C}[1]{>{\centering\arraybackslash}p{#1}}
\newcolumntype{M}[1]{>{\centering\arraybackslash}m{#1}}
\allowdisplaybreaks

\def\_#1{{\bf #1}}

\def\.{\cdot}

\def\Re{{\rm Re\mit}}
\def\Im{{\rm Im\mit}}
\def\l#1{\label{eq:#1}}
\def\r#1{(\ref{eq:#1})}

\newcommand{\ve}[1]{\mathbf{#1}}

\def\H0{{H_0}}
\def\E0{\eta_0 {H_0}}

\def\w{\omega}

\def\=#1{\overline{\overline #1}}

\begin{document}

\title{Parametric Mie resonances and directional amplification in time-modulated scatterers}

%


\author{V.~Asadchy$^{1,2,\dagger}$}
\thanks{These authors contributed equally to this work}
\author{A.G.~Lamprianidis$^{3}$}
\thanks{These authors contributed equally to this work}
\author{G.~Ptitcyn$^{2}$}
\author{M.~Albooyeh$^{4}$}
\author{Rituraj$^{1,5}$}
\author{T.~Karamanos$^{3}$}
\author{R.~Alaee$^3$}
\author{S.A.~Tretyakov$^2$}
\author{C.~Rockstuhl$^{3,6}$}
\author{S.~Fan$^1$}

\affiliation{
$^1$Ginzton Laboratory and Department of Electrical Engineering, Stanford University, Stanford, California 94305, USA\\
$^2$Department of Electronics and Nanoengineering, Aalto University, P.O.~Box 15500, FI-00076 Aalto, Finland\\
$^3$Institute of Theoretical Solid State Physics, Karlsruhe Institute of Technology, 76131 Karlsruhe, Germany\\
$^4$Mobix Labs Inc., 15420 Laguna Canyon, Irvine, California 92618, USA\\
$^5$Department of Electrical Engineering, Indian Institute of Technology Kanpur, Kanpur-208016, UP, India\\
$^6$Institute of Nanotechnology, Karlsruhe Institute of Technology, 76131 Karlsruhe, Germany\\
$^\dagger$email: viktar.asadchy@aalto.fi \hspace{4cm}}

\begin{abstract}
We provide a theoretical description of  light scattering by a spherical particle  whose permittivity is modulated in time at twice the frequency of the incident light. 
Such a particle acts as a finite-sized photonic time crystal and, despite its sub-wavelength spatial extent, can host optical  parametric amplification. Conditions of parametric Mie resonances in the sphere are derived. We show that      time-modulated materials  provide a   route to tailor directional light amplification, qualitatively different from that  in scatterers made from a gain media. We design two characteristic time-modulated spheres that simultaneously exhibit   light amplification and desired radiation patterns, including those with zero backward and/or vanishing forward scattering. The latter sphere    provides an opportunity for creating  shadow-free detectors of incident light.
\end{abstract}

\maketitle   

\section{Introduction}
Sub-wavelength high-index dielectric resonators provide a versatile platform for light control at the nanoscale.   These resonators can support strong light localization described by  multipolar Mie-type resonances~\cite{mishchenko2002scattering,bohren_absorption_2008,kruk_functional_2017,liu_generalized_2018,tzarouchis_light_2018,koshelev_dielectric_2021}. The resonant modes are generated by the volumetric distribution of displacement currents and can be of   electric or  magnetic kinds. A remarkable feature of Mie-type scattering lays in the possibility to spectrally overlap   several multipolar modes for engineering complex scattering patterns. 
During the last few years, such multipolar mode engineering led to  a number of applications in nanophotonics, including  wavefront manipulations for metasurfaces~\cite{decker_high-efficiency_2015}, bound states in the continuum~\cite{hsu_bound_2016,koshelev_asymmetric_2018},    nonradiating anapole modes~\cite{devaney_radiating_1973,miroshnichenko_nonradiating_2015}, nanoparticle localization~\cite{bag_transverse_2018}, and  directional spontaneous parametric down-conversion~\cite{marino_spontaneous_2019,nikolaeva_directional_2021}, among many others.

Most of the previous works on Mie-type scatterers concentrated on time-invariant particles whose permittivity \textit{does not} change in time. 
The time variation of material properties unlocks an additional dimension of control  in electromagnetic systems~\cite{engheta_metamaterials_2021,galiffi_photonics_2022}. Recently, a wide range of novel optical effects was suggested based on time-varying materials, such as 
photonic time crystals~\cite{biancalana_dynamics_2007,zurita-sanchez_reflection_2009,reyes-ayona_observation_2015,lustig_topological_2018,park_spatiotemporal_2021,sharabi2021disordered}, temporal discontinuities~\cite{zhou2020broadband,pacheco-pena_antireflection_2020,quinones_tunable_2021,yin_efficient_2022},  time-varying meta-atoms and antennas~\cite{salandrino_plasmonic_2018,mirmoosa_timevarying_2021,solis_functional_2021,mekawy_parametric_2021,mirmoosa_dipole_2022}, effective magnetic field for photons~\cite{fang_realizing_2012}, optically induced negative refraction~\cite{vezzoli_optical_2018}, synthetic dimensions~\cite{yuan_synthetic_2021}, etc. 
The  temporal material   modulation has the potential to dramatically extend both conceptual and applied aspects of Mie-type scattering~\cite{stefanou_light_2021,ptitcyn_scattering_2021}.  However, to date this area of research has remained essentially unexplored. 

In this work, we analyse  light scattering by a sphere  whose permittivity is modulated at twice the frequency of the incident light, which corresponds to the case of parametric excitation.  Based on Floquet-Mie theory and the temporal coupled mode theory, we demonstrate  that  such a sphere, despite its sub-wavelength spatial extend,   hosts  parametric Mie resonances.  
It is revealed that temporal modulations provide an additional design dimension, allowing directional light amplification by a scatterer. We highlight a qualitative difference of this mechanism from   light amplification in scatterers with gain.
We design two characteristic examples of parametric scatterers possessing finite light amplification with desired scattering patterns.
A related effect of parametric amplification in spherical scatterers with the second-order nonlinearity was recently reported in Ref.~\cite{jahani_wavelength-scale_2021}, however,  simultaneous far-field pattern engineering  was not demonstrated. 

\section{Bulk time-modulated  medium}
\begin{figure}[tb]
	\centering
\includegraphics[width=0.99\linewidth]{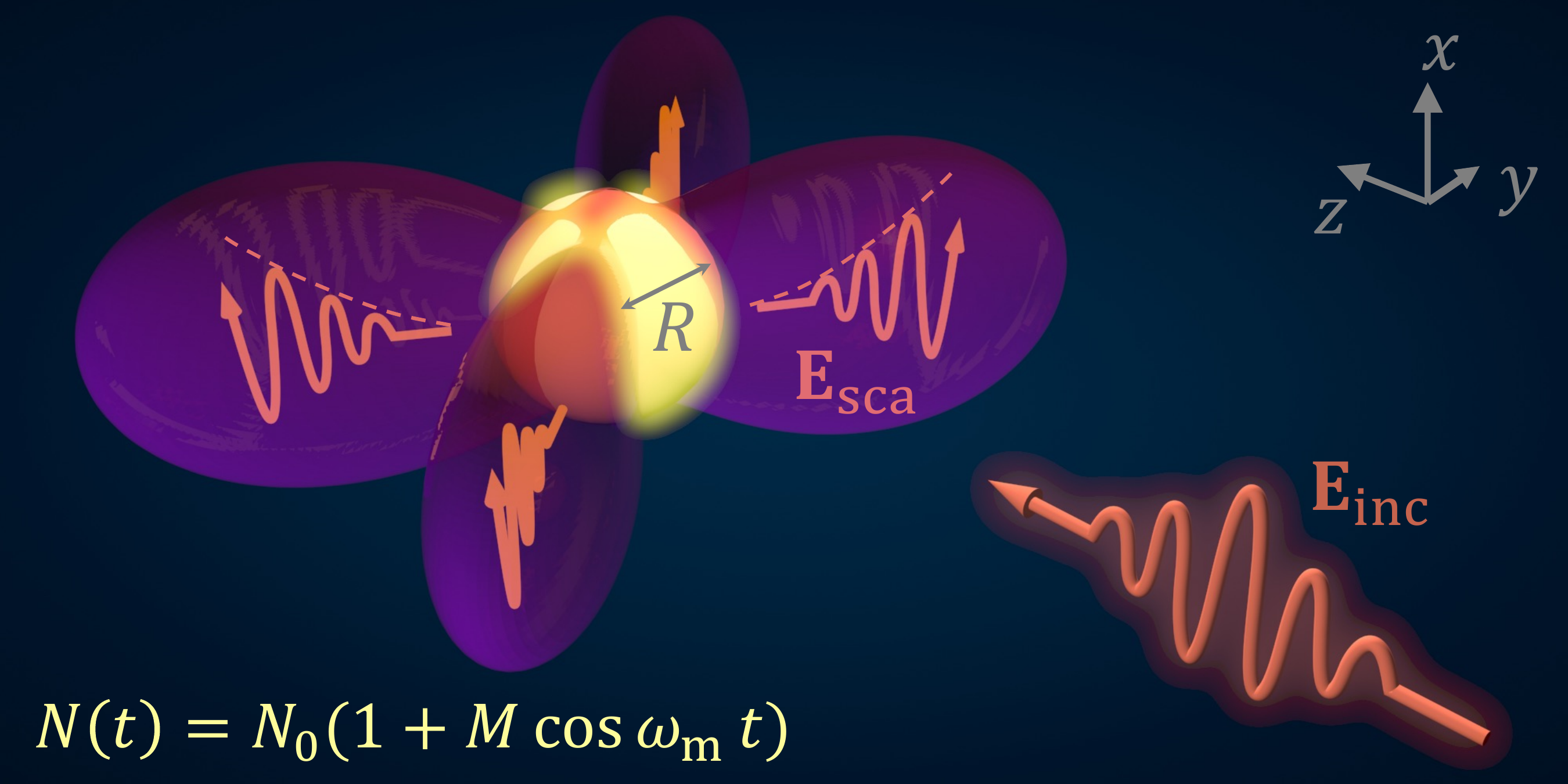} 
	\caption{Spherical particle with time-modulated bulk carrier density  illuminated by incident light. Temporal modulation leads to parametric Mie resonances with simultaneous scattered-field amplification and possibility of far-field pattern manipulation. }
	\label{fig1}
\end{figure}

We consider a sphere located at the center of the coordinate system (see Fig.~\ref{fig1}).
The material of the sphere without modulations is described by a single-pole Lorentz-Drude dispersion model with the  \textit{sta\-tiona\-ry} relative permittivity function given by $\varepsilon_{\rm st}(\w)=1+\w_{\rm p}^2/ (\w_{\rm r}^2-\w^2-i \gamma  \w)$, where  $\gamma $ is the damping factor and $\w_{\rm r}$   the   resonance frequency.
In what follows, we choose without loss of generality a 
plasma frequency of $\w_{\rm p}= \sqrt{N_0 q_{\rm e}^2/m_{\rm e} \varepsilon_0}=3.5\w_{\rm r}$, where $q_{\rm e}$ and $m_{\rm e}$ are the electron charge and mass, respectively, and $\varepsilon_0$ is the vacuum permittivity.  
Parameter $N_0$ is the time-averaged bulk carrier density.
The temporal variation of the sphere's  permittivity  $\varepsilon$  is assumed to be via the modulation of the charge carrier density of the form $N(t)=N_0 (1+M\cos \w_{\rm m}t)$ (see Sec.~1 of the Supplemental Material~\cite{suppl}), where  $M$ is the modulation strength  and $\w_{\rm m}$ is the modulation frequency. In what follows, we choose a regime of relatively low dispersion, that is, $\w_{\rm m}=0.5 \w_{\rm r}$.  Modulation of the carrier concentration  with the strength of the order of unity and $\w_{\rm m}$ at   optical frequencies
was  experimentally demonstrated in several recent works~\cite{alam_large_2016,caspani_enhanced_2016,vezzoli_optical_2018}.

We first find the eigenfrequencies and corresponding eigenmodes of an unbounded dispersive material with time-varying carrier concentration $N(t)$.
The wave equation of such material written for the electric field $\ve{E}(\mathbf{r},\omega)$ reads~\cite{mirmoosa_dipole_2022,shi_multi-frequency_2016,ptitcyn_scattering_2021}
\begin{eqnarray}
    \nabla\times\nabla\times \mathbf{E}(\mathbf{r},\omega)\hspace{165pt}&&\nonumber\\
    ={k}^2(\omega)\left[\mathbf{E}(\mathbf{r},\omega)+\int\limits_{-\infty}^{+\infty}\chi (\omega-\omega',\omega')\mathbf{E}(\mathbf{r},\omega')\mathrm{d}\omega'\right].\hspace{10pt}&& \l{eq1}
\end{eqnarray}
Here, $k(\w)=\w/c$ is the wavenumber of free space, $c$ is the speed of light, $\ve{r}$ is the position vector,   $\chi (\omega-\omega',\omega') = \varepsilon (\omega-\omega',\omega') - \delta (\omega-\omega')$ is the generalized susceptibility that describes  the polarization density   at frequency $\w$ induced by an electric field harmonic at frequency $\w'$, and $\delta (\w -\w')$ is the Dirac delta function. 
This susceptibility incorporates   the information about the dynamics of the modulated medium and its dispersion properties~\cite{solis_time-varying_2021,mirmoosa_dipole_2022}.
Solving  the wave equation, we look for the electric
field  in the form 
$ \ve{E}(\mathbf{r},\omega)=\int A(\kappa)S_\kappa(\omega) \mathbf{F}(\kappa  \_r)\mathrm{d}\kappa $, where  
$A(\kappa)$ is the complex modal amplitude, and
$S_\kappa(\omega)$ and 
$\mathbf{F}(\kappa \_r)$  are the spectral and spatial  parts of the eigenmodes, respectively~\cite{ptitcyn_scattering_2021}. The latter is a solution of the Helmholtz wave equation with eigen-wavenumber $\kappa$.

By substituting  the electric field ansatz into~\r{eq1}, we obtain the following eigenvalue equation in the matrix form (see Sec.~1 of the Supplemental Material~\cite{suppl}): 
\begin{equation}
     k_n^2 (\varepsilon_{{\rm st},n} S_{\kappa,n} + \varepsilon_{{\rm dyn},n} S_{\kappa,n+1} + \varepsilon_{{\rm dyn},n} S_{\kappa,n-1} ) = \kappa^2 S_{\kappa,n},
      \l{eq2}
\end{equation}
where $\varepsilon_{\rm dyn}(\w)=[\varepsilon_{\rm st}(\w)-1] M/2$ is the \textit{dynamic} part of the relative permittivity.
In~\r{eq2}, index $n$ means that the corresponding function is taken at frequency $\w_n=\w + n \w_{\rm m}$. 

\begin{figure}[tb]
	\centering
\includegraphics[width=0.9\linewidth]{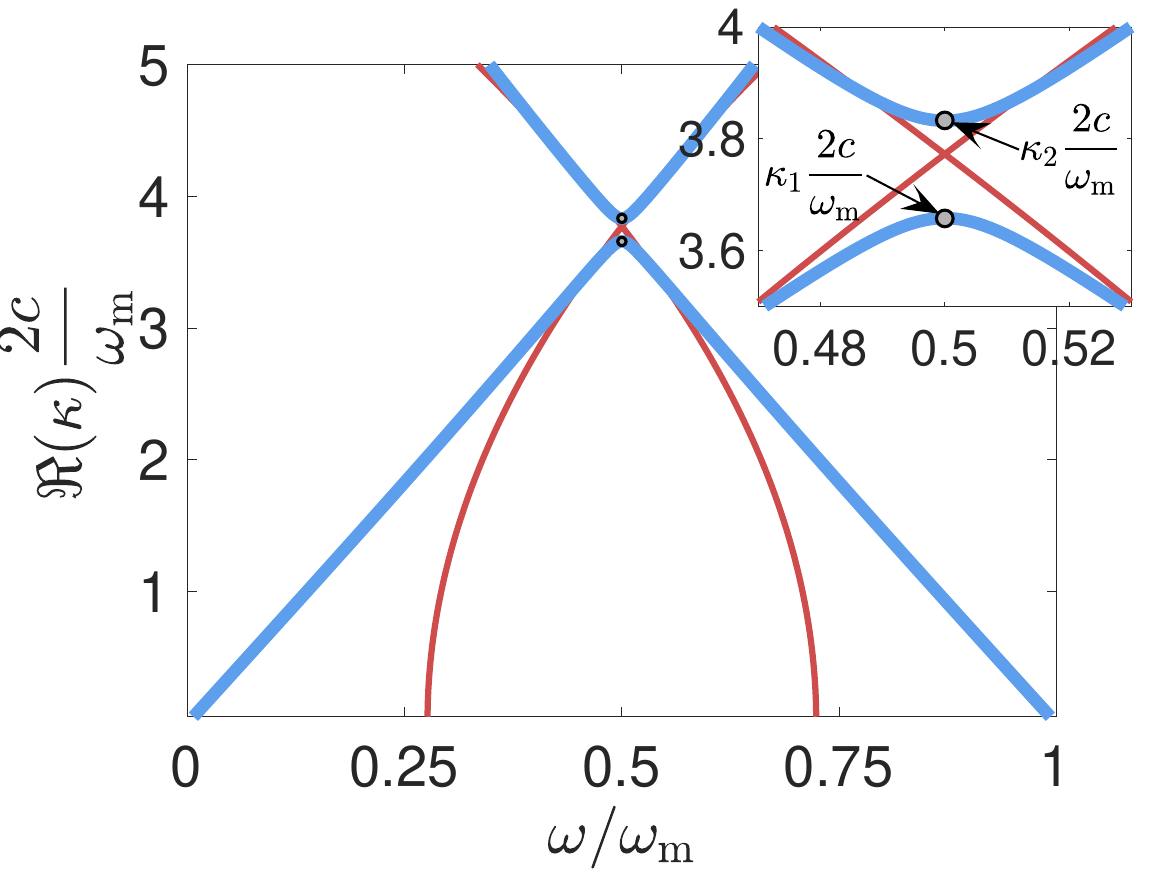} 
	\caption{Band structure diagram of a time-modulated material plotted for the case with modulation strength~$M=0.1$ and damping factor $\gamma=0$~Hz. The thick blue and thin red lines correspond, respectively, to low- and high-frequency  bulk plasmon-polariton bands in the Lorentzian dispersion. }
	\label{fig2}
\end{figure}

Equation~\r{eq2} allows one to find a set of eigen-wavenumbers $\kappa_q$ ($q$ is a positive integer) for a bulk temporally modulated  material at a given Floquet frequency $\w$~\cite{zurita-sanchez_reflection_2009}, as well as  the  matrix of weights $S_{q n}$ of the modes  with frequency $\w_n$ and    wavenumber $\kappa_q$.
Eigenvalue equation~\r{eq2} results in a   band diagram with period $\omega_{\rm m}$ that corresponds to that of a photonic time crystal. 
Such a band diagram  is dual (under replacement $\kappa \leftrightarrow \w$) to that of conventional photonic crystals~\cite{joannopoulos_photonic_2011}. 
According to the duality with conventional photonic crystals,  photonic time crystals can host momentum bandgaps. 
By solving eigenvalue equation~\r{eq2} numerically, we are able to plot in Fig.~\ref{fig2} a band diagram of our photonic time crystal for the special case of the material with $M=0.1$ and $\gamma=0$~Hz (presence of a small nonzero $\gamma$ leads to additional bands in the diagram but does not significantly modify the dispersion within the gap).

Since the considered material has a Lorentzian dispersion, there are two bulk plasmon-polariton bands where the real part of the permittivity is positive. These two bands are shown with blue and red lines in the figure. The first one (blue) is split by a  momentum bandgap, inside which  there are two modes which have purely imaginary eigenfrequencies (one attenuating and one  amplifying)~\cite[p.~53]{joannopoulos_photonic_2011,lustig_topological_2018}. 
The amplifying mode is responsible for the parametric amplification effect in time-modulated materials and it is being excited  even if the bandgap is closed by the red bands. The effect of the red bands in the scattering by the sphere can be neglected. Note that parametric amplification   should be distinguished from  optical  gain that is modeled by a negative damping factor~$\gamma$. 

\section{Parametric Mie resonances in time-modulated spheres}
Next, we analyse wave phenomena in a finite-size sphere made from  a time-modulated material.  
For clarity of the analysis, here we assume that temporal modulation inside the sphere are uniform. As we show in Sec.~3 of the Supplemental Material~\cite{suppl}, possible spatial inhomogeneities of the sphere has only a minor quantitative impact on the results. 
First, we  find   the condition of optical parametric amplification.  For its derivation, we will consider a separate eigenvalue problem for the electric field amplitudes across the sphere boundary with no incident field (parametric    oscillations). 
To find the parametric oscillation condition  analytically, we   consider the Floquet frequency right at the center  of the momentum bandgap, that is, $\w =\w_{\rm m}/2$, and
exploit the weak-modulation approximation~\cite{martinez-romero_parametric_2018}, which works perfectly in the regime of  $M \ll 1$ and provides a very satisfactory estimation for $M<0.2$ (see Fig.~S2 in Sec.~5 of the Supplemental Material~\cite{suppl}). Here, we apply the approximation solely for the sake of making theoretical analysis more transparent for the reader and highlighting the qualitative picture of the considered phenomena. It is important to mention that one can also solve the eigenvalue equation~\r{eq2} exactly, without resorting to any approximations, which will be done for the scatterer examples considered below.
As we verified numerically, under this approximation, there are only two dominant harmonics $ \w_0= \w_{\rm m}/2$ and $\w_{-1}= -\w_{\rm m}/2$ and two dominant (lowest) momentum bands $\kappa_1$ and $\kappa_2$. In other words, the matrix of modal weights $S_{q n}$ can be truncated to merely a $2\times 2$ size with indices $q=\{1,2\}$ and $n=\{0,-1\}$. The points with $\kappa_1$ and $\kappa_2$ are marked in the diagram of Fig.~\ref{fig2}. Using the approximation, equation~\r{eq2} can be solved analytically in a closed form (see Sec.~2 of the Supplemental Material~\cite{suppl}) yielding the following expressions for the momenta and modal weights for the parametric-oscillation regime:
\begin{equation} 
\begin{array}{cc}\displaystyle
   \kappa_1 = \frac{\w_{\rm m}}{2c} \sqrt{\Re{\,\varepsilon_{{\rm st},0} - \tilde{\varepsilon}}}, \quad
     \kappa_2 = \frac{\w_{\rm m}}{2c} \sqrt{\Re{\,\varepsilon_{{\rm st},0} + \tilde{\varepsilon}}},  & \vspace{0.3cm}  \\  \displaystyle 
  S_{q n} =  
  \left(\begin{array}{cc}
\varepsilon_{{\rm dyn},0}^*/(i \Im{\, \varepsilon_{{\rm st},0}}  - \tilde{\varepsilon}) & \qquad 1  \vspace*{.4cm}\\
\varepsilon_{{\rm dyn},0}^*/(i \Im{\, \varepsilon_{{\rm st},0}}  + \tilde{\varepsilon}) & \qquad 1  
\end{array}\right), 
  & 
\end{array}
      \l{eq3}
\end{equation}
where  ``$*$'' denotes complex conjugation and $\tilde{\varepsilon}=\sqrt{| \varepsilon_{{\rm dyn},0} |^2 -(\Im{\,  \varepsilon_{{\rm st},0}})^2} $.
For the case  when $\gamma=0$~Hz, the matrix   simplifies into $S_{q n}=[-1, 1; 1, 1]$ and the momentum bandgap width $\Delta \kappa=\kappa_2-\kappa_1$ is linearly proportional to the modulation amplitude $M$:
\begin{equation} 
\Delta \kappa= M \, \frac{\w_{\rm m}}{4c} \, \frac{\w_{\rm p}^2}{\sqrt{\w_{\rm r}^2-\w_{\rm m}^2/4} \sqrt{\w_{\rm r}^2-\w_{\rm m}^2/4+\w_{\rm p}^2}}.
      \l{bandgap001}
\end{equation}


%
%
Due to the spherical symmetry, the  electric field inside the   sphere   can be expressed using  a set of  VSHs as $\displaystyle \ve{E}^{\rm in}(\ve{r},\w_n)= \sum_{\alpha,\mu,\nu,q} A^{\rm in}_{\alpha \mu \nu q} \, \ve{F}^{(1)}_{\alpha \mu \nu} (\kappa_q \_r) S_{qn}$, with $A^{\rm in}$ standing for   amplitudes of corresponding VSHs with wavenumber~$\kappa_q$. Here, indices  $\mu$ and  $\nu$ stand for the angular momentum along the $z$-axis and  the multipolar order, respectively~\cite{ptitcyn_scattering_2021},\cite[Sect.~13.3]{morse1953methods}.
Subscript $\alpha$    stands for one of the two labels, $\alpha_M$ or $\alpha_N$, and    refers to magnetic   or  electric multipolar modes, respectively. Finally, superscript $\iota$  takes the values ``1'' or ``3'' to refer to regular or radiating VSHs, respectively.  
The electric field outside the sphere (in vacuum), represented by the scattered field only, is given by 
$\displaystyle \ve{E}^{\rm sca}(\ve{r},\w_n)= \sum_{\alpha,\mu,\nu} A^{\rm sca}_{\alpha \mu \nu} (\w_n) \ve{F}^{(3)}_{\alpha \mu \nu} (k_n \_r) $. Importantly, here we are looking for the solution with no incident field present, which corresponds to the parametric oscillations regime.
Next, we substitute these expressions  into the boundary conditions at the surface of the sphere with radius~$R$ ($\ve{r}=R \, \hat{\ve{r}}$)~\cite{ptitcyn_scattering_2021}
\begin{equation} 
\begin{array}{cc}\displaystyle
\hat{\ve{r}} \times \left[ \ve{E}^{\rm in}(\hat{\ve{r}} R,\w_n) - \ve{E}^{\rm sca}(\hat{\ve{r}} R,\w_n) \right] =0, \vspace{0.3cm} \\ \displaystyle
\hat{\ve{r}} \times \left[ \ve{H}^{\rm in}(\hat{\ve{r}} R,\w_n) - \ve{H}^{\rm sca}(\hat{\ve{r}} R,\w_n) \right] =0,
\end{array}
      \l{S15}
\end{equation}
where $\hat{\ve{r}} $ is the radial unit vector and $R$ is the radius of the sphere. 
Using the orthogonality relations for vector spherical harmonics~\cite{ptitcyn_scattering_2021}, we obtain the following system of equations:
\begin{equation} 
\begin{array}{cc}\displaystyle
\sum_{q=1}^2  A^{\rm in}_{\alpha \mu \nu q} \, S_{qn} z_{\alpha \nu}^{(1)} (\kappa_q R) =
A^{\rm sca}_{\alpha \mu \nu} (\w_n) z_{\alpha \nu}^{(3)} (k_n R),   \vspace{0.3cm} \\ \displaystyle
\hspace{-0.2cm}
\sum_{q=1}^2  A^{\rm in}_{\alpha \mu \nu q} \, S_{qn} \kappa_q z_{\beta \nu}^{(1)} (\kappa_q R) =
A^{\rm sca}_{\alpha \mu \nu} (\w_n) k_n z_{\beta \nu}^{(3)} (k_n R).
\end{array}
      \l{eq4}
\end{equation}
Here, index $\beta$   is always different from $\alpha$, that is, if $\alpha=\alpha_M$ then $\beta=\alpha_N$, and vice versa. Function $z_{\alpha_M \nu}^{(\iota)} $ denotes the spherical Bessel ($\iota=1$) and Hankel ($\iota=3$) functions of the first kind of order~$\nu$, while  $z_{\alpha_N \nu}^{(\iota)}(x)=\frac{1}{x}\frac{\partial}{\partial x}[x z_{\alpha_M \nu}^{(\iota)}(x)]$.
Equations~\r{eq4} must hold for each set of parameters $\{\alpha, \mu,\nu,n\}$. Writing these two equations for the two frequency harmonics $n=0$ and $n=-1$, we finally formulate the eigenvalue equation for
the electric field amplitudes across the sphere boundary, i.e., with respect to field amplitudes $A^{\rm in}_{\alpha \mu \nu 1}$, $A^{\rm in}_{\alpha \mu \nu 2}$, $A^{\rm sca}_{\alpha \mu \nu} (\w_{-1})$, and $A^{\rm sca}_{\alpha \mu \nu} (\w_{0})$. 
For the regime of parametric oscillations in the sphere (in the absence of incident waves), we are looking for the  solutions with nonzero amplitudes  $A^{\rm in}$ and $A^{\rm sca}$. Therefore, we equate the determinant of the $4\times4$ matrix in the eigenvalue problem to zero and solve the resulting equation with respect to the radius~$R$ and the modulation strength~$M$ of the sphere (see Sec.~2 of the Supplemental Material~\cite{suppl}).

Figure~\ref{fig3} depicts with colored lines the solutions of the zero matrix determinant for electric-type ($\alpha=\alpha_N$) and magnetic-type  ($\alpha=\alpha_M$)  modes in the sphere with multipolar orders from $\nu=1$ to $\nu=5$, indicating the threshold values of the modulation strength to provide parametric oscillations. The data are plotted for   $\gamma=0$~Hz. Non-zero dissipation would lead to merely a minor change in Fig.~\ref{fig3}, shifting all the curves to the upper side. The solutions are independent of parameter~$\mu$.
The   lines in the figure show all the sets of  parameters ($R$ and $M$) which yield parametric amplification of the corresponding multipolar mode in the time-modulated sphere.
 \begin{figure}[tb]
	\centering
\includegraphics[width=0.99\linewidth]{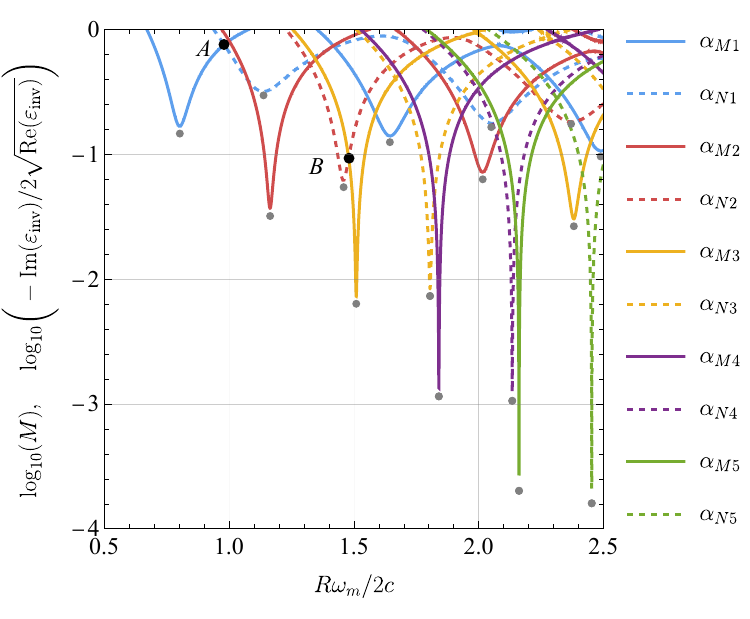} 
	\caption{
	The colored curves  depict threshold values of the modulation strength $M$ that provide parametric oscillations  at fixed frequency 
	$\omega_{\rm m}/2$ for different multipolar modes  in a time-modulated sphere 
	versus its normalized radius. The grey dots depict values of the normalized imaginary part of permittivity that support lasing  at fixed frequency $\omega_{\rm las}= \omega_{\rm m}/2$ for different   modes  in a time-invariant sphere with optical gain. 
	While   horizontal coordinates of these points  match to those of the minima of the colored curves    for corresponding multipolar modes, their vertical coordinates do differ and the difference depends on the value of  chosen stationary permittivity $\varepsilon_{\rm st,0} = \Re (\varepsilon_{\rm inv})$ (see Sec.~4 of Supplemental Material~\cite{suppl}). 
	}
	\label{fig3}
\end{figure}
One can observe from the plot that higher-order multipolar modes (with larger values of $R\w_{\rm m}/2c$ and higher quality factors) can host parametric oscillations at   lower values of~$M$. 
For example, the magnetic multipole  of the order $\nu=5$ ($\alpha=\alpha_{M5}$, green solid line) exhibits parametric oscillation at the value of~$M$ as low as $2.27 \times 10^{-4}$. 
The normalized radii $R\w_{\rm m}/2c$ at the  dips in Fig.~\ref{fig3} approximately coincide with those of  conventional Mie resonances $R\w_{\rm Mie}/c$ of the corresponding modes in a  non-modulated sphere.

To analyse  the physics of parametric Mie resonances, we employ a temporal coupled-mode theory~\cite{haus_waves_1984,haus_electromagnetic_2012,rodriguez_2_2007,fan_12_2008}.
Let us consider two coupled quasi-normal~\cite{doost_resonantstate_2014,PhysRevA.101.011803,PhysRevB.102.075103} modes inside the sphere at frequencies $\pm \w_{\rm m}/2$  with the total electric field of the form
$ \ve{E} (\ve{r},t) =  a_1(t) \, {\rm e}^{-i \w_{\rm m} t/2} \, \_E_{\rm Mie} (\ve{r}) + a_2(t) \, {\rm e}^{-i \w_{\rm m} t/2} \, [\_E_{\rm Mie} (\ve{r})]^*  + {\rm c.c.}$
Here, $a_1(t)$ and $a_2(t)$  are the  slowly varying   temporal   envelopes of the original and time-reversed modes and $\_E_{\rm Mie} (\ve{r})$ is the spatial mode profile.
We assume that $\w_{\rm m}/2$ is close to the frequency $\w_{\rm Mie}$ that corresponds to   one of the stationary Mie resonances, that is, $\w_{\rm Mie} =\w_{\rm m}/2 -\Delta \w -i \gamma_{\rm tot}$ (where $|\Delta \w + i\gamma_{\rm tot}| \ll  \w_{\rm m}/2 $). Here, $\gamma_{\rm tot}$ is the total decay rate which includes radiation and possible dissipation losses (due to positive $\gamma$). 
Starting from the wave equation in the time-modulated material, one can arrive to the following system of coupled-mode equations describing   evolution of mode envelopes $a_1(t)$ and $a_2^*(t)$ inside the sphere (see Sec.~3 of the Supplemental Material~\cite{suppl}): 
\begin{equation} 
\begin{array}{cc}\displaystyle
 \frac{{\rm d} }{{\rm d} t} a_1(t) = \left[ i \Delta \w - \gamma_{\rm tot} \right] a_1(t)   + i\eta a_2^*(t),
 \vspace{0.1cm} \\ \displaystyle
\frac{{\rm d} }{{\rm d} t} a_2^*(t) = \left[- i \Delta \w - \gamma_{\rm tot} \right] a_2^*(t)   - i\eta^* a_1(t),
 \end{array}
      \l{eq5}
\end{equation}
where $\eta$ is a coupling parameter   linearly proportional to modulation strength~$M$. 
Solving   system~\r{eq5}, we obtain the   threshold value of modulation strength $M_{\rm thr} \propto \gamma_{\rm tot} + \frac{1}{2\gamma_{\rm tot}} \Delta \w^2 $
for parametric amplification in the sphere.
This value provides a qualitative description of the spectral lineshapes of the parametric Mie resonances (note  that in  Fig.~\ref{fig3} the logarithm of $M$ is plotted). 
For   modes with higher multipolar orders~$\nu$,  the decay rate due to radiation loss~$\gamma_{\rm tot}$ is smaller, which results in   deeper dips.

As is seen from Fig.~\ref{fig3}, the curves depicting the parametric oscillation condition at fixed frequency
$\omega_{\rm m}/2$ are continuous. This feature allows us to select the sphere configuration  with~$M$ and~$R$ at the points where the curves intersect such that simultaneous parametric amplification  of two desired multipolar   modes occurs at the same frequency (ensuring coherence). The orientation of these modes is locked when the sphere is illuminated by  incident light. By choosing the pair of modes, one can control the radiation pattern of the amplified scattered light.
Importantly, such a multi-mode coherent amplification regime is not accessible in time-invariant spheres made from a medium with gain~\cite{olmos-trigo_kerker_2020}.   In order to demonstrate this, we additionally mark with grey dots in Fig.~\ref{fig3}  those   configurations  of such an active sphere  (with radius $R$ and complex time-invariant permittivity $\varepsilon_{\rm inv}$)  that support lasing  (divergent scattering cross section) for different   modes at the fixed frequency $\omega_{\rm las}= \omega_{\rm m}/2$.
For fair comparison, we choose   $\Re (\varepsilon_{\rm inv})=\varepsilon_{\rm st,0}$. 
 The details of the calculations as well as comparison for other values of $\Re (\varepsilon_{\rm inv})$ can be found in Sec.~4 of Supplemental Material~\cite{suppl}. 
As is seen, lasing in time-invariant spheres occurs only at discrete points in the configuration space, and simultaneous satisfaction of the lasing condition for several modes at the same frequency is generally impossible.  
Such qualitatively different behaviour suggests that
temporal modulations   provide    a  pathway for achieving  coherent  amplification by the sphere with desired radiation pattern.
Moreover, 
due to a finite width of each dip in   Fig.~\ref{fig3}, it is possible to excite  higher-order multipolar modes in a sphere of smaller size compared to that   in the absence of temporal modulations~\cite{salandrino_plasmonic_2018}.

\section{Scattering from time-modulated spheres}
In order to demonstrate the potential of directional amplification, next we consider two representative examples of parametric spheres. 
In both  examples
the sphere is illuminated by   monochromatic plane waves at a frequency $\w_{\rm inc}$ (see Fig.~\ref{fig1}). The incident frequency is slightly shifted away from $\w_{\rm m}/2$ so that we can achieve finite and controllable amplification and use the harmonic-field analysis. 
From a practical point of view, the amplification can be locked-in to frequency $\w_{\rm inc}$ instead of $\w_{\rm m}/2$ if temporal modulations occur while the sphere is illuminated by the incident light~\cite{boyd_nonlinear_2020}.  

Designing the radiation pattern of a  particle near the lasing condition (near parametric oscillation) is challenging. Whereas the lasing   occurs for each multipole independently, we need to obtain the superposition of multipoles of comparable strength and with appropriate phases  to achieve a desired radiation pattern. However, the lasing multipoles have diverging amplitudes and therefore dominate the radiation pattern, rendering the contributions of the rest of the radiating multipoles insignificant upon a superposition. Therefore, the simultaneous satisfaction of the lasing condition for several multipoles is needed to shape the radiation pattern of a lasing particle. Fine tuning the system at the vicinity of the parameter space, where such an overlap of parametric Mie resonances happens,  allows for the engineering of the relative amplitudes and phases of each lasing multipole, finally leading to the engineering of a lasing particle with a desired radiation pattern.

For the first example, we consider a sphere configuration with $M=0.68$ and $R=1.048 \frac{2c}{\w_{\rm m}}$, marked by point~A in Fig.~\ref{fig3}. The configuration corresponds to the first parametric resonance crossing of the electric and magnetic dipole modes. Since contours  in  Fig.~\ref{fig3} were plotted under the approximation of $M \ll 1$, for finding the exact coordinates of point~A, we calculated the contours considering  a large number of frequency harmonics (see Sec.~5 in Supplemental Material~\cite{suppl}).  
 In the present and the following examples, we chose $\gamma=0$~Hz. We excite the sphere by incident light at $\w_{\rm inc}=0.498 \w_{\rm m}$. 
To find the scattered fields, we use the eigenvalue equation~\r{eq2},  the  boundary conditions which include the incident fields, and the expansion of the fields in series of radiating VSHs (see Sec.~6 in Supplemental Material~\cite{suppl}).
Figure~\ref{fig4}(a) depicts the scattered far-field pattern at frequency $\w_{\rm inc}$. The pattern  is unidirectional, revealing zero backward scattering due to close fulfillment of the first Kerker condition~\cite{kerker_electromagnetic_1983}. The condition implies that the electric and magnetic modes in the sphere have approximately same amplitudes and phases. We were able to reach such a balance by fine adjustments of parameters $M$, $R$, and $\w_{\rm inc}$.
Interestingly, we observed that having  a non-zero damping factor $\gamma$ in the  material of the sphere  precludes achieving exact zero backward scattering, which is in agreement with recent similar findings for time-invariant  lossy  uniform spheres~\cite{olmos-trigo_kerker_2020,olmos-trigo_optimal_2020}. While in~\cite{olmos-trigo_kerker_2020} it was proved that ideal zero backward scattering cannot occur in spheres with \textit{optical gain}, this statement does not apply to the time-modulated spheres with \textit{parametric gain} considered in this work. 
 \begin{figure}[tb]
	\centering
\includegraphics[width=0.95\linewidth]{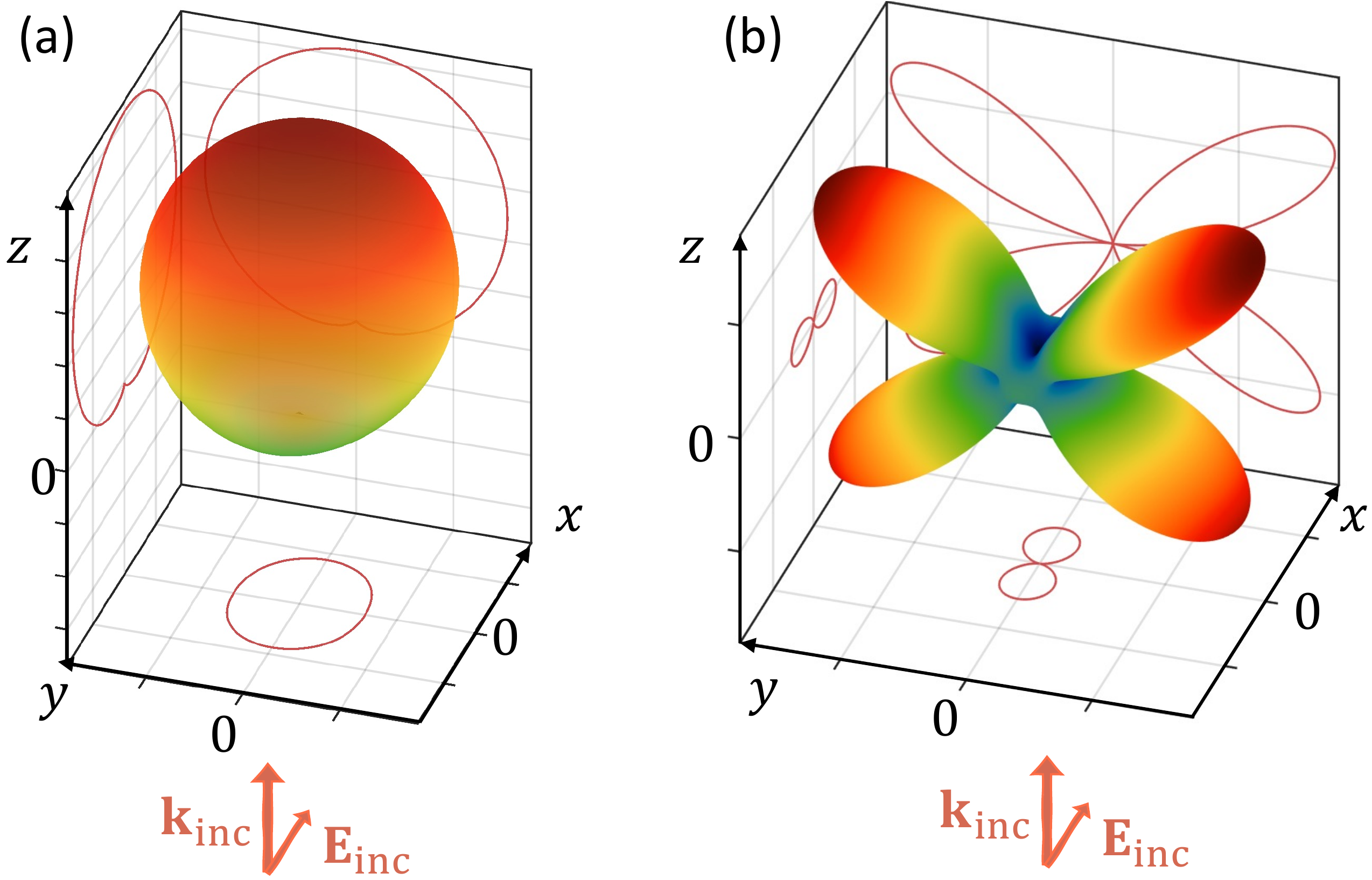} 
	\caption{Scattered far-field patterns of time-modulated spheres with parameters (a) $M=0.68$ and $R=1.048 \frac{2c}{\w_{\rm m}}$ and (b)  $M=0.093$ and $R=1.481 \frac{2c}{\w_{\rm m}}$. The patterns are calculated at the frequency of incident wave~$\w_{\rm inc}$. The red contours depict  cross sections of the   patterns parallel to the $xy$, $yz$, and $xz$ planes, calculated at the center of the coordinate system. The colors of the patterns denote the scattering amplitude (dark red  and dark blue colors stand for the maximum and  minimum values, respectively).
	}
	\label{fig4}
\end{figure}
The scattering and absorption cross sections in this example are $C_{\rm sca}/C_{\rm geom} =2629.2$ and $C_{\rm abs}/C_{\rm geom} = -2627.5$, where $C_{\rm geom}= \pi R^2$  and negative sign of $C_{\rm abs}$ implies the activity of the modulated sphere.
Clearly, the scattering cross section largely exceeds that of the same sphere without temporal modulations (for which case $C_{\rm sca}^{\rm st}/C_{\rm geom} =5.5$ and $C_{\rm abs}^{\rm st}=0$) due to the presence of   modulation.

The second example is a sphere with a configuration of $M=0.093$ and $R=1.481 \frac{2c}{\w_{\rm m}}$  (see point~B in Fig.~\ref{fig3})  which coincides with the parametric resonance crossing of the electric quadrupole and magnetic octupole modes.  
Incident light at   $\w_{\rm inc}=0.4995 \w_{\rm m}$ is scattered by the sphere with the pattern shown in Fig.~\ref{fig4}(b). The pattern has   sharp dips in both the backward and  forward  directions.  Note that, whereas the electric and magnetic dipoles have opposite parity symmetry, ensuring the first Kerker condition, the electric quadrupole and magnetic octupole have the same parity symmetry, allowing for the engineering of both the first and second Kerker conditions simultaneously~\cite{kerker_electromagnetic_1983}.
The scattering and absorption cross sections are
$C_{\rm sca}/C_{\rm geom} = 858.3$ and $C_{\rm abs}/C_{\rm geom} = -857.5$ (in comparison, $C_{\rm sca}^{\rm st}/C_{\rm geom} = 2.53$ and $C_{\rm abs}^{\rm st}=0$ for the stationary sphere).
For both considered time-modulated  spheres, the   optical theorem~\cite{newton_optical_1976}, written  for  the forward scattering and   extinction cross section at the  fundamental frequency $\w_{\rm inc}$, is  satisfied.
The peculiar pattern in Fig.~\ref{fig4}(b) with scattering dips in both forward and backward directions stems from the precise engineering of amplitude and phases of the two multipolar modes (see Sec.~6 of the Supplemental Material~\cite{suppl} and~\cite{lee_simultaneously_2018}). 

\section{Discussion}
We have explored optical parametric amplification by  spherical scatterers with time-modulated permittivity. 
The presented  two example  geometries highlight the fascinating opportunities of simultaneous 
light amplification and scattering pattern control provided by the additional temporal dimension. 
Indeed, the second   sphere example   provides
an interesting functionality: shadow-free detection of incident light due to vanishing forward scattering (related concept using  active and parity-time-symmetric dimers     was suggested in~\cite{fleury_invisible_2015,safari_shadowfree_2018a}). 
The sphere scatters   light sideways where it can be detected by sensors. Parametric amplification enables      detection of   extremely weak   signals. 
Due to the  symmetry of the sphere, it is possible to determine also the propagation direction  of the   light under detection by looking at the  scattering pattern.
Furthermore, time-modulated  particles can find applications for designing  nanoscale    amplifiers. Due to the directional nature of their scattering and possibility of finite amplification,  one can create exotic non-attenuating waveguide modes and topological edge modes in  a non-uniform lattice of such spheres.   Our results can be extended to other  domains (acoustics, water waves, etc.), to particles with other geometries, and represent  the first step towards parametric metasurfaces   based on  time-modulated scatterers.

\section*{Acknowledgment}
This work was supported by the MURI
project from the U.S. Air Force of Office of Scientific Research (Grant No FA9550-21-1-0244), 
German Research Foundation through Germany’s Excellence Strategy via
the Excellence Cluster 3D Matter Made to Order (EXC-
2082/1 - 390761711), 
Alexander von Humboldt Foundation, and the Academy of Finland (project 330260). A. G. L. acknowledges support
from the Max Planck School of Photonics, which is supported by BMBF, Max Planck Society, and Fraunhofer
Society and from the Karlsruhe School of Optics and
Photonics (KSOP).



\bibliography{references}
\end{document}